\documentclass{cimento}

\usepackage{graphicx}  

\title{Mass Spectra of $\Xi$ and $\Omega$ Baryons using hypercentral Constituent Quark Model}
\author{Chandni Menapara\thanks{chandni.menapara@gmail.com}
	\atque
	Ajay Kumar Rai}
\instlist{\item {Department of Physics, Sardar Vallabhbhai National Institute of Technology, Surat-395007, Gujarat, INDIA.}}

\begin{document}
	
	\maketitle
	
	\begin{abstract}
		Hadron spectroscopy is an important tool towards the study of internal quark dynamics in a composite system. The present article focuses on the study of resonance spectra of strange baryons with S=-2, -3. The non-relativistic approach utilizes screened potential as hypercentral one to obtain masses. The spin-dependent part for all possible hyperfine states has been incorporated. 
	\end{abstract}

	\section{Introduction}
	Numerous fundamental questions related to the nature of matter and its structure are brought up by strong interaction physics. The primary goals of hadronic physics are to identify the pertinent degrees of freedom that govern hadronic phenomena at all scales and then establish the connection of these degrees of freedom to the parameters and fundamental fields of QCD \cite{qcd}. The light, strange sector encompasses a lot of yet unknown or missing resonances leading to scarce experimental data \cite{crede, thiel}.\\
	With these motivation, the current article briefly shows the hypercentral Constituent Quark Model (hCQM) been employed to obtain resonance masses. The later section highlights linear Regge trajectories for the obtained results.\\
	$\Omega$ and $\Xi$ as both are not easily observed in experiments and the little knowledge relies on the bubble chamber data\cite{pdg}. Algebraic models have been explored in ways to understand internal baryon dynamics, wherein U(7) has been used in a recent work \cite{amiri}. R. Bijker and group too have focused on algebraic approach with the string-like model \cite{bijker}. The quark-diquark model has been studied in various versions since long back; Gursey Radicati-inspired exchange interaction employed by E. Santopinto et al to reproduce all strange and non-strange baryon resonances \cite{s5}. Also, the relativistic study with quark-diquark model have been studied for all sectors \cite{barabanov}. Regge phenomenology has also been employed in the study of light, strange baryons using n and J plane linear curves \cite{juhi}. Other than these various models have been employed over the years \cite{chen,klempt}.

	\section{Method: Hypercentral Constituent Quark Model (hCQM)}
	
	Constituent Quark models (CQM) recognize a baryon to be three-quark system confined by interaction, wherein constituent quark mass taking care of internal effects. One of the choice of hypercentral potential form is screened potential \cite{amee1,amee3}. Earlier the results have been studied for linear form \cite{cpc1,cpc3,ijmpa1,akram}. 
	The hyper-radial equation is as shown below\cite{gianinni},
	\begin{equation}
		\left[\frac{d^{2}}{dx^{2}} + \frac{5}{x}\frac{d}{dx} - \frac{\gamma(\gamma +4)}{x^{2}}\right]\psi(x) = -2m[E-V_{3q}(x)]\psi(x) 
	\end{equation}
	The confining part of the potential appears as,
	\begin{equation}
		V^{0}(x)=a\left(\frac{1-e^{-{\mu} x}}{\mu}\right)
	\end{equation}
	Such potential has been known to show good results using hCQM for heavy quark system including mesons and baryons \cite{zalak19,zalak18,zalak16,keval18,keval20}. The screening parameter is different in case of heavy and light systems. Based on earlier studies, the screening parameter $\mu$ has been varied over a range and 0.3 has been considered as the value obtain the spectra for the systems considered here \cite{raghav}. \\
	
	The spin-dependent interaction to take care of hyperfine splitting needs to be incorporated separately as $V_{SD}(x)$ with spin-orbit, spin-spin, and tensor terms \cite{voloshin}.
	\begin{eqnarray}
		V_{SD}(x) = V_{LS}(x)({\bf L \cdot S}) + V_{SS}(x) \left[S(S+1)-\frac{3}{2}\right] \\ + V_{T}(x)\left[S(S+1) -\frac{3({\bf S \cdot x})({\bf S \cdot x})}{x^{2}} \right] 
	\end{eqnarray}
	Let, $V_{V}= \frac{\tau}{x} $ and $V_{S}= \alpha x $.\\
	With this non-relativistic approach, a number of resonances have been obtained. 
	
	\section{Mass Spectra}
	\label{sec3}
	$\Xi$ baryons appear with isospin $I=\frac{1}{2}$ in the octet ($J=\frac{1}{2}$) and decuplet ($J=\frac{3}{2}$) as $\Xi$ and $\Xi^{*}$ respectively. The quark combination is uss for $\Xi^{0}$ and dss for $\Xi^{-}$. The $\Omega$ baryon is completely associated with the decuplet representation which is the least explored in light, strange sector. With increase in strangeness, the resonances get scarcely observed. Many experiments are focused towards this goal namely, BESIII, PANDA, etc. \cite{panda2,panda7}  \\
	Here, table \ref{tab:Xi-screen} denotes that 2S states with $J^P = \frac{3}{2}^{+}$ is differing by 21 MeV from PDG. Also, the 1P$\frac{3}{2}^{-}$, 2P$\frac{1}{2}^{-}$ and 1D$\frac{5}{2}^{+}$ are well in accordance with experimental known masses within variation of 50 MeV. \\
	Table \ref{tab:omega-screen} signifies the limited $\Omega$ baryon states. The 1P$\frac{1}{2}^{-}$, 2P$\frac{1}{2}^{-}$ and 1D$\frac{3}{2}^{+}$ with 2012, 2380 and 2250 respectively are quite near to obtained results. So, for lower excited states, screened potential with a fixed screening parameter reproduces the observed results. However, the spin splitting as compared to our earlier works of linear potential decreases. 
	\begin{table}[h]

		\begin{minipage}{0.48\textwidth}
			\renewcommand{\arraystretch}{1.2}
			\centering
			\caption{\label{tab:Xi-screen} $\Xi$ resonance mass spectra using Screened Potential (in MeV)}
			\begin{tabular}{ccccc}
				\hline
				State & $J^{P}$ & $M_{scr}$ & $M_{exp}$\cite{pdg}\\
				\hline
				1S & $\frac{1}{2}^{+}$ & 1321 & 1321\\
				& $\frac{3}{2}^{+}$ & 1531 & 1532 \\
				2S & $\frac{1}{2}^{+}$ & 1842\\
				& $\frac{3}{2}^{+}$ & 1971 & 1950\\
				3S & $\frac{1}{2}^{+}$ & 2320 & 2370 \\
				& $\frac{3}{2}^{+}$  & 2442\\
				4S & $\frac{1}{2}^{+}$ & 2888\\
				& $\frac{3}{2}^{+}$ & 2997\\ 
				
				\hline
				$1^{2}P_{1/2}$ & $\frac{1}{2}^{-}$ & 1877 & \\
				$1^{2}P_{3/2}$ & $\frac{3}{2}^{-}$ & 1869 & 1823\\
				$1^{4}P_{1/2}$ & $\frac{1}{2}^{-}$ & 1882 & \\
				$1^{4}P_{3/2}$ & $\frac{3}{2}^{-}$ & 1873 & 1823\\
				$1^{4}P_{5/2}$ & $\frac{5}{2}^{-}$ & 1862 & \\
				\hline
				$2^{2}P_{1/2}$ & $\frac{1}{2}^{-}$ & 2340 & 2370*\\
				$2^{2}P_{3/2}$ & $\frac{3}{2}^{-}$ & 2328 & \\
				$2^{4}P_{1/2}$ & $\frac{1}{2}^{-}$ & 2347\\
				$2^{4}P_{3/2}$ & $\frac{3}{2}^{-}$ & 2334\\
				$2^{4}P_{5/2}$ & $\frac{5}{2}^{-}$ & 2318\\
				\hline
				$1^{2}D_{3/2}$ & $\frac{3}{2}^{+}$ & 2265\\
				$1^{2}D_{5/2}$ & $\frac{5}{2}^{+}$ & 2246 & 2250\\
				$1^{4}D_{1/2}$ & $\frac{1}{2}^{+}$ & 2287\\
				$1^{4}D_{3/2}$ & $\frac{3}{2}^{+}$ & 2273\\
				$1^{4}D_{5/2}$ & $\frac{5}{2}^{+}$ & 2253\\
				$1^{4}D_{7/2}$ & $\frac{7}{2}^{+}$ & 2229\\
				\hline
				$2^{2}D_{3/2}$ & $\frac{3}{2}^{+}$ & 2802\\
				$2^{2}D_{5/2}$ & $\frac{5}{2}^{+}$ & 2778\\
				$2^{4}D_{1/2}$ & $\frac{1}{2}^{+}$ & 2828\\
				$2^{4}D_{3/2}$ & $\frac{3}{2}^{+}$ & 2810\\
				$2^{4}D_{5/2}$ & $\frac{5}{2}^{+}$ & 2787\\
				$2^{4}D_{7/2}$ & $\frac{7}{2}^{+}$ & 2758\\
				\hline
				$1^{2}F_{5/2}$ & $\frac{5}{2}^{-}$ & 2717 \\
				$1^{2}F_{7/2}$ & $\frac{7}{2}^{-}$ & 2679\\
				$1^{4}F_{3/2}$ & $\frac{3}{2}^{-}$ & 2760\\
				$1^{4}F_{5/2}$ & $\frac{5}{2}^{-}$ & 2729\\
				$1^{4}F_{7/2}$ & $\frac{7}{2}^{-}$ & 2690\\
				$1^{4}F_{9/2}$ & $\frac{9}{2}^{-}$ & 2644\\
				\hline
				$2^{2}F_{5/2}$ & $\frac{5}{2}^{-}$ & 3321\\
				$2^{2}F_{7/2}$ & $\frac{7}{2}^{-}$ & 3277\\
				$2^{4}F_{3/2}$ & $\frac{3}{2}^{-}$ & 3369\\
				$2^{4}F_{5/2}$ & $\frac{5}{2}^{-}$ & 3340\\
				$2^{4}F_{7/2}$ & $\frac{7}{2}^{-}$ & 3290\\
				$2^{4}F_{9/2}$ & $\frac{9}{2}^{-}$ & 3238\\
				\hline
			\end{tabular}
		\end{minipage}
		\hfill
		\begin{minipage}{0.48\textwidth}
			\renewcommand{\arraystretch}{1.2}
			\centering
			\caption{\label{tab:omega-screen} $\Omega$ resonance mass spectra using Screened Potential (in MeV)}
			\begin{tabular}{ccccc}
				\hline
				State & $J^{P}$ & $M_{scr}$ & $M_{exp}$ \cite{pdg}\\
				\hline
				1S & $\frac{3}{2}^{+}$ & 1672 & 1672\\
				2S & $\frac{3}{2}^{+}$ & 2063\\
				3S & $\frac{3}{2}^{+}$ & 2428\\
				4S & $\frac{3}{2}^{+}$ & 2839\\ 
				5S & $\frac{3}{2}^{+}$ & 3290\\
				\hline
				$1^{2}P_{1/2}$ & $\frac{1}{2}^{-}$ & 1998 & 2012\\
				$1^{2}P_{3/2}$ & $\frac{3}{2}^{-}$ & 1993\\
				$1^{4}P_{1/2}$ & $\frac{1}{2}^{-}$ & 2001\\
				$1^{4}P_{3/2}$ & $\frac{3}{2}^{-}$ & 1996\\
				$1^{4}P_{5/2}$ & $\frac{5}{2}^{-}$ & 1989\\
				\hline
				$2^{2}P_{1/2}$ & $\frac{1}{2}^{-}$ & 2352 & 2380\\
				$2^{2}P_{3/2}$ & $\frac{3}{2}^{-}$ & 2345\\
				$2^{4}P_{1/2}$ & $\frac{1}{2}^{-}$ & 2356\\
				$2^{4}P_{3/2}$ & $\frac{3}{2}^{-}$ & 2349\\
				$2^{4}P_{5/2}$ & $\frac{5}{2}^{-}$ & 2339\\
				\hline
				
				$1^{2}D_{3/2}$ & $\frac{3}{2}^{+}$ & 2289 & 2250\\
				$1^{2}D_{5/2}$ & $\frac{5}{2}^{+}$ & 2278\\
				$1^{4}D_{1/2}$ & $\frac{1}{2}^{+}$ & 2301\\
				$1^{4}D_{3/2}$ & $\frac{3}{2}^{+}$ & 2293\\
				$1^{4}D_{5/2}$ & $\frac{5}{2}^{+}$ & 2282\\
				$1^{4}D_{7/2}$ & $\frac{7}{2}^{+}$ & 2269\\
				\hline
				$2^{2}D_{3/2}$ & $\frac{3}{2}^{+}$ & 2684\\
				$2^{2}D_{5/2}$ & $\frac{5}{2}^{+}$ & 2671\\
				$2^{4}D_{1/2}$ & $\frac{1}{2}^{+}$ & 2698\\
				$2^{4}D_{3/2}$ & $\frac{3}{2}^{+}$ & 2689\\
				$2^{4}D_{5/2}$ & $\frac{5}{2}^{+}$ & 2676\\
				$2^{4}D_{7/2}$ & $\frac{7}{2}^{+}$ & 2660\\
				\hline
				
				$1^{2}F_{5/2}$ & $\frac{5}{2}^{-}$ & 2614 \\
				$1^{2}F_{7/2}$ & $\frac{7}{2}^{-}$ & 2594\\
				$1^{4}F_{3/2}$ & $\frac{3}{2}^{-}$ & 2636\\
				$1^{4}F_{5/2}$ & $\frac{5}{2}^{-}$ & 2620\\
				$1^{4}F_{7/2}$ & $\frac{7}{2}^{-}$ & 2600\\
				$1^{4}F_{9/2}$ & $\frac{9}{2}^{-}$ & 2576\\
				\hline
				$2^{2}F_{5/2}$ & $\frac{5}{2}^{-}$ & 3044\\
				$2^{2}F_{7/2}$ & $\frac{7}{2}^{-}$ & 3022\\
				$2^{4}F_{3/2}$ & $\frac{3}{2}^{-}$ & 3067\\
				$2^{4}F_{5/2}$ & $\frac{5}{2}^{-}$ & 3050\\
				$2^{4}F_{7/2}$ & $\frac{7}{2}^{-}$ & 3029\\
				$2^{4}F_{9/2}$ & $\frac{9}{2}^{-}$ & 3003\\
				\hline
				
			\end{tabular}
		\end{minipage}
	\end{table}
	Some of the few one- and two-star states with spin-parity are discussed in detail, and a tentative place in either orbital or radial state is assigned to them. It should be noted that the screening is not as effective for lower states, but that screening leads to a decrease in mass for higher excited states and a significant decrease in hyperfine splitting. More experimental results shall be guiding force towards establishing the complete spectra. 
	
	\section{Regge Trajectories}
	One of the helpful tools in spectroscopic research has been Regge trajectories. The square of mass relates linearly to total angular momentum J and principal quantum number n which actually is a part of Chew-Frautschi plots \cite{regge}. \begin{equation}
		J = aM^{2} + a_{0} \quad \quad
		n = b M^{2} + b_{0}
	\end{equation}

	
	\begin{figure}
		\begin{minipage}{0.48\textwidth}
			\includegraphics[width=0.9\textwidth]{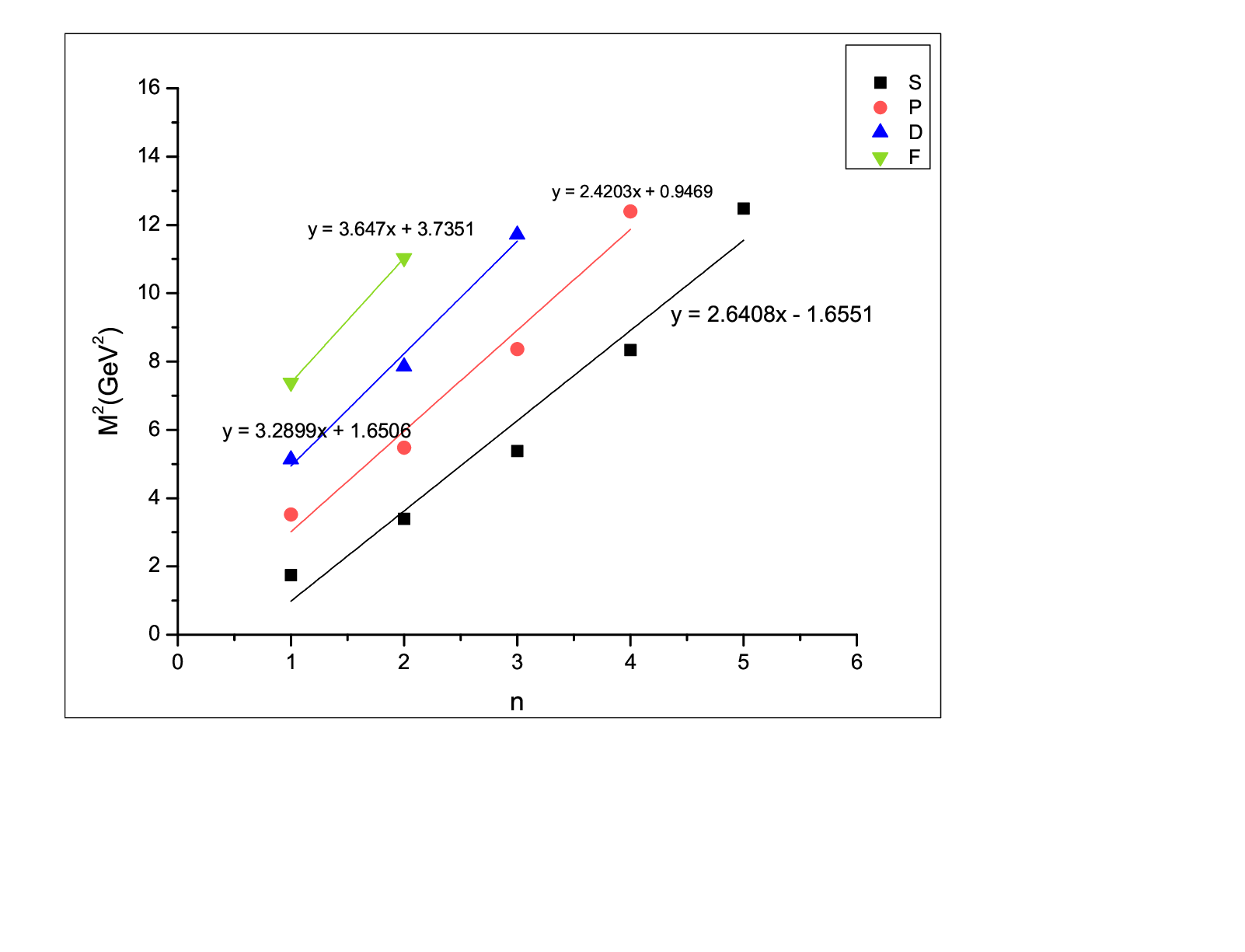} 
			\caption{\label{fig:xi-nm-s} Regge trajectory $\Xi$ for $n \rightarrow M^{2}$ for screened potential.}	
		\end{minipage}
		\hfill
		\begin{minipage}{0.48\textwidth}
			\includegraphics[width=0.9\textwidth]{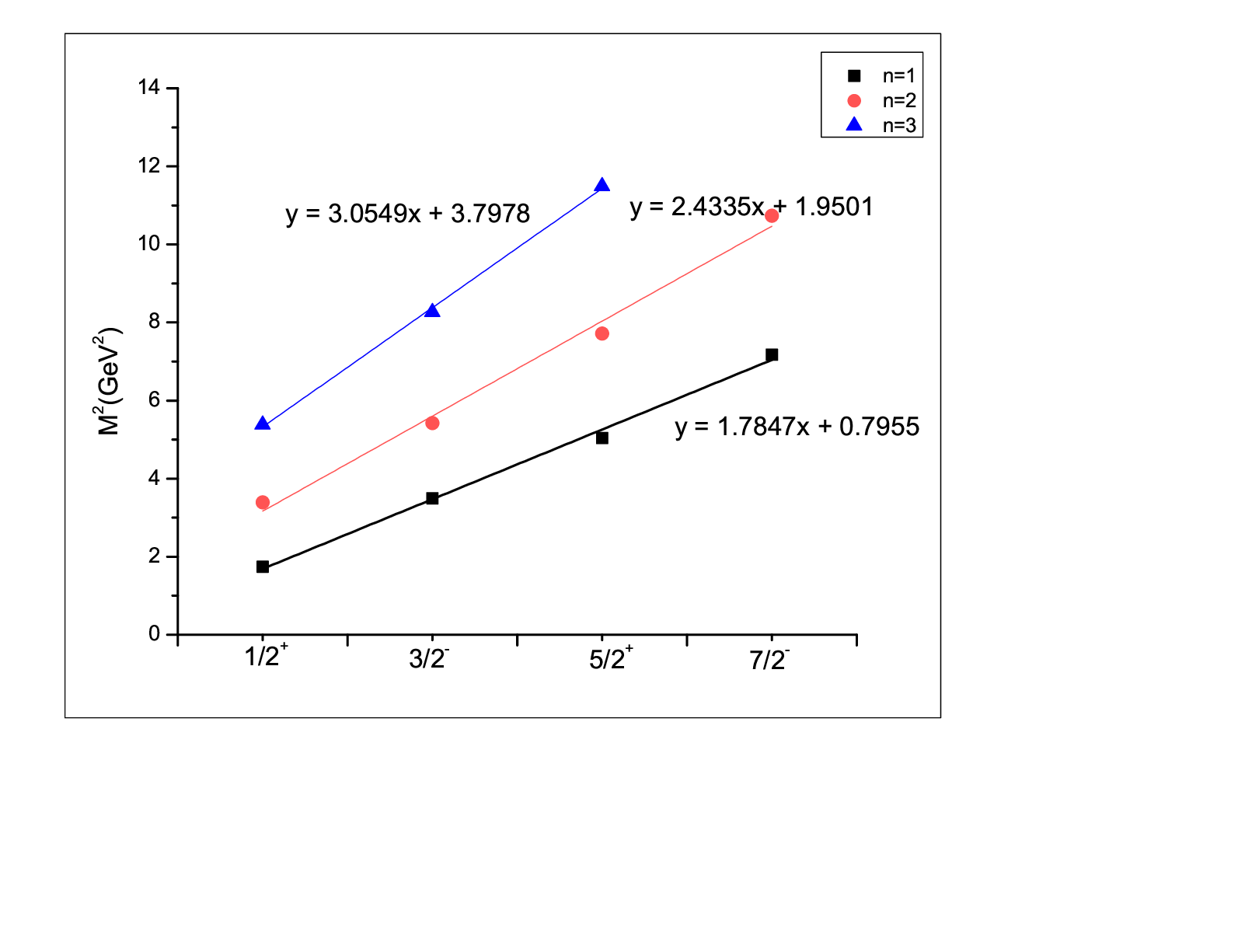}
			\caption{\label{fig:xi-j1-s} Regge trajectory $\Xi$ for $J^{P} \rightarrow M^{2}$ for screened potential.}
		\end{minipage}
	\end{figure}
	
	\begin{figure}
		\begin{minipage}{0.48\textwidth}
			\includegraphics[width=0.9\textwidth]{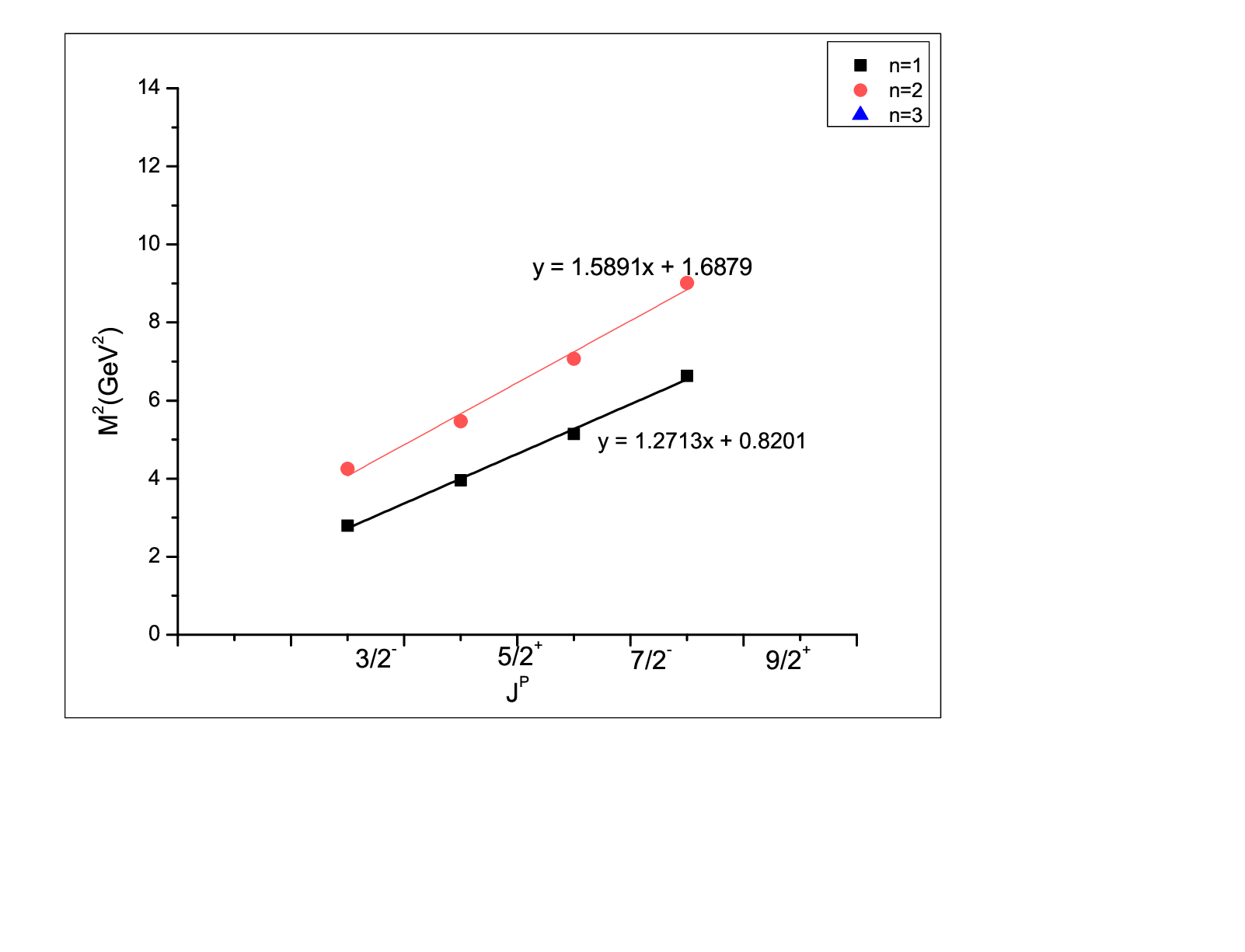} 
			\caption{\label{fig:o-nm-s} Regge trajectory $\Omega$ for $n \rightarrow M^{2}$ for screened potential.}	
		\end{minipage}
		\hfill
		\begin{minipage}{0.48\textwidth}
			\includegraphics[width=0.9\textwidth]{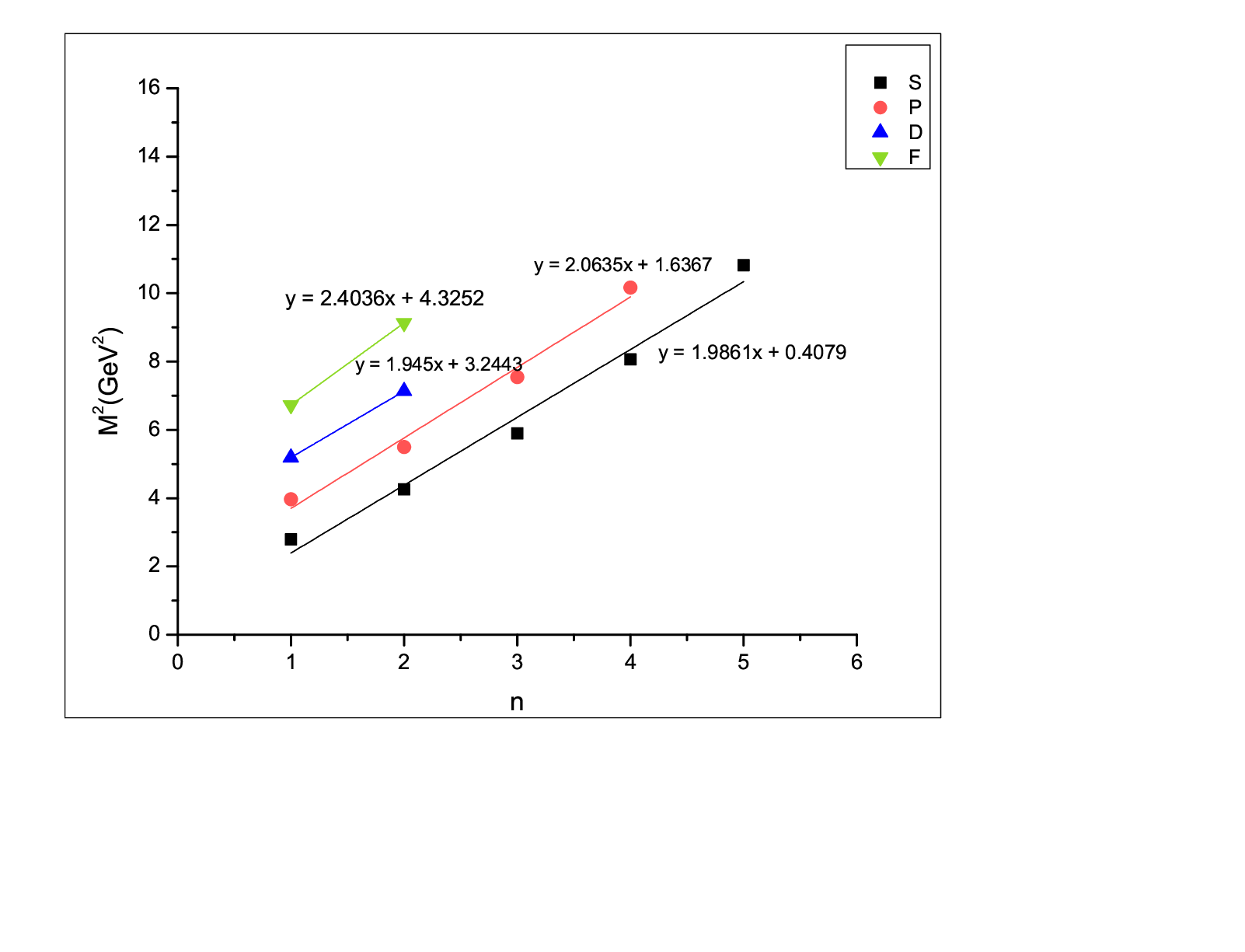}
			\caption{\label{fig:o-j1-s} Regge trajectory $\Omega$ for $J^{P} \rightarrow M^{2}$ for screened potential.}
		\end{minipage}
	\end{figure}
	Figures \ref{fig:xi-nm-s}, \ref{fig:xi-j1-s}, \ref{fig:o-nm-s} and \ref{fig:o-j1-s} show the linearly fitted mass obtained from the model here. However, due to scarcity of available experimental data, exact position of any given mass cannot be commented. But new mass obtained through experiments can be placed on these lines to predict their possible spin-parity assignment as well. 
	\section{Conclusion and Outlook}
	The non-relativistic hCQM has been a tool to obtain resonances of strange baryons where experimental observations are very few. An attempt has been made to identify the effect of screened potential on light, strange spectra. The low-lying excited states are well within the range of Particle Data Group (PDG). The screening effect shall be more significant at higher excited states. The results' tables shows the possible spin-parity assignment of some one-star states especially for $\Omega$ baryon. $\Omega$(2012) is a state whose precise structure is debated between three-quark or molecular type. This study allows us to look for predicted range of states in experiments and study different properties.
	
	\acknowledgments
	The authors are thankful to the organizers of HADRON2023, Genoa, Italy for the opportunity to present our work. Ms. Chandni Menapara acknowledges the support from DST under INSPIRE FELLOWSHIP.


\begin{thebibliography}{0}
		\bibitem{qcd}\BY{Gross F. \textsc {\it et al.}} {\it Eur. Phys. J. C} {\bf 83} (2023) 1125 
		\bibitem{crede}\BY{Crede V. \atque Roberts W.} {\it Rept. Prog. Phys.} {\bf 76} (2013) 076301  
		\bibitem{thiel}\BY{Thiel A., Afzal F. \atque Wunderlich Y.} {\it Prog. Part. Nucl. Phys.} {\bf 125} (2022) 103949   
		\bibitem{pdg}\BY{Workman R. L. \textsc {\it et al.}} [Particle Data Group], {\it Prog. Theor. Exp. Phys.} {\bf 2022} (2022)  083C01 
		\bibitem{amiri}\BY{Amiri N., Ghapanvari M. \atque Jafarizadeh M. A.} {\it Eur. Phys. J. Plus} {\bf 141} (2021) 136  
		\bibitem{bijker}\BY{Bijker R. \textsc {\it et al.}} {\it Phys. Rev. D.} {\bf 94} (2016) 074040 
		\bibitem{s5}\BY{Santopinto E.} {\it Phys. Rev. C} {\bf 72} (2005) 022201; {\it Phys. Rev. C} {\bf 92} (2015) 025202 
		\bibitem{barabanov}\BY{Barabanov B.Yu.} {\it Prog. Part. Nucl. Phys.} {\bf 116} (2021) 103835 
		\bibitem{juhi}\BY{Oudichhya J.,  Gandhi K. \atque  Rai A. K.} {\it Nucl. Phys. A} {\bf 1035} (2023) 122658 
		\bibitem{chen}\BY{Chen Y. \atque  Ma B. Q.} {\it Chin. Phys. Lett.} {\bf 25} (2008) 3920  
		\bibitem{klempt}\BY{Klempt E.} {\it Phys. Rev. C} {\bf 66} (2002) 058201 
		
		\bibitem{amee1}\BY{Kakadiya A., Shah Z., Gandhi K. \atque Rai A. K.} {\it Few-Body Systems} {\bf 63} (2022)29; {\it Int. J. Mod. Phys. A} {\bf 37} (2022) 2250053 
		\bibitem{amee3}\BY{Kakadiya A., Menapara C. \atque Rai A. K.} {\it Int. J. Mod. Phys. A} {\bf 38} (2023) 234103 
		\bibitem{cpc1}\BY{Menapara C., Shah Z. \atque Rai A. K.} {\it Chin. Phys. C} {\bf 45} (2021) 023102 
		\bibitem{cpc3}\BY{Menapara C. \atque Rai A. K.} {\it Chin. Phys. C} {\bf 46} (2022) 103102; {\bf 45} (2021) 063108 
		\bibitem{ijmpa1}\BY{Menapara C. \atque Rai A. K.} {\it Int. Jour. Mod. Phys. A} {\bf 37} (2022) 2250177; {\bf 38} (2023) 2350053 
		\bibitem{akram}\BY{Ansari A., Menapara C. \atque Rai A. K.} {\it Int. J. Mod. Phys. A} {\bf 38} (2023) 2350108
		\bibitem{gianinni}\BY{Giannini M. M. \atque Santopinto E.} {\it Chin. J. Phys.} {\bf 53} (2015) 020301
		\bibitem{zalak19}\BY{Shah Z. \textsc {\it et al.}} {\it Chin. Phys. C} {\bf 43} (2019) 034102 
		\bibitem{zalak18}\BY{Shah Z. \atque Rai A. K.} {\it Few-Body Syst} {\bf 59} (2018) 112; {\it Eur. Phys. J. A} {\bf 53} (2017)  195  
		\bibitem{zalak16}\BY{Shah Z.,  Thakkar K. \atque Rai A. K.} {\it Eur. Phys. J. C} {\bf 76} (2016)  530
		\bibitem{keval18}\BY{Gandhi K., Shah Z. \atque Rai A. K.} {\it Eur. Phys. J. Plus} {\bf 133} (2018) 1  
		\bibitem{keval20}\BY{Gandhi K. \atque Rai A. K.} {\it Eur. Phys. J. Plus} {\bf 135} (2020) 213 
		
		
		\bibitem{raghav}\BY{Chaturvedi R. \atque Rai A. K.} {\it Int. J. Theo. Phys.} {\bf 59} (2020) 3508 
		\bibitem{voloshin}\BY{Voloshin M. B.} {\it Prog. Part. Nucl. Phys.} {\bf 61} (2008) 455 
		\bibitem{panda2}\BY{Barruca G. \textsc {\it et al.}} [PANDA Collaboration] {\it Eur. Phys. J. A} {\bf 57} (2021) 184 
		\bibitem{panda7}\BY{Abazov V. \textsc {\it et al.}} [PANDA Collaboration] arXiv:2304.11977
		\bibitem{regge}\BY{Tang A. \atque Norbury J.} {\it Phys. Rev. D} {\bf 62} (2000) 016006
		
	\end{thebibliography}
\end{document}